# Influence of rotation speed on natural frequency: A short introduction and presentation of an imaginary 'antigravity' world

Christopher Provatidis

*Department of Mechanical Engineering, National Technical University of Athens, Zografou Campus, 9 Heroes of Polytechnion Ave., 157 80 Athens, GREECE*

We present an instructive way to introduce the subject of resonance using a typical spring-mass system without damping. In contrast to the standard approach found in most textbooks, here we propose a way to extend the lectures from common translational systems to rotating ones. In the latter case, particular attention is paid to the role of the variable centripetal force. The students will find analytical solutions of the simplest possible ordinary differential equations of motion and see that the natural frequency strongly depends on the rotation speed, the latter being a simple vector sum. The analysis reveals that always the resultant ground force is harmonic. In order to depict the significance in the influence of rotation on natural frequency, we present an imaginary 'antigravity' world in which the aforementioned dependency is absent.

## I. INTRODUCTION

There are lots of resonance phenomena in physics and particularly in mechanics. The easiest test case is a spring-mass system without damping. The spring, of stiffness $k$, is fixed on the still ground, while its other end is attached to a mass $m$; a force $F(t)$ is exerted on the mass. Newton's second law is applied and the ordinary differential equation of motion is derived. In order to continue with the introduction of the subject, the students must be aware (or we have to remind them) how to solve an ordinary differential equation (ODE). The analytical solution appears a denominator that when becoming equal to zero it causes an infinite displacement at the resonance.[1]

Although the above knowledge is well absorbed by all physicists and engineers, not many are aware about the dependency of the natural frequency on the possible rotation of the physical system. Of course there are some relevant research works in physics but they are very specialized and of rather complicated character.[2] Also, in mechanical engineering there is a lot of rotating machinery such as turbines, rolling bearings, gears, machining tools etc. Following a standard handbook, "*if the natural frequency of the structure of an equipment coincides with the frequency of the applied vibration, the vibration condition may be made much worse as a result of resonance*".[3] Despite the fact that this is standard knowledge, a fast gallop conducted by the author has revealed that on this topic there exists a gap of knowledge in the majority of the students. As not a simple way to introduce this subject exists in textbooks, we were motivated to write this paper.

As a side effect, we also found an opportunity to please the readers by discussing some issues of an interesting topic that has not been solved during the last three centuries (after Sir Isaac Newton ceased): antigravity. The readers will enjoy that wrong assumptions, based on the independency of the natural frequency on rotation, can lead to the wrong conclusion that a harmonic vibration or a constant internal force may lead to non-harmonic ground forces, which would be permanently upward or downward; in other words, antigravity impulse terms of the form $\alpha t$ (linear in time $t$) would appear.

## II. RESONANCE

Let us assume a spring-mass ($k$, $m$) system on which a force $F(t)$ is exerted. For the sake of briefness, we consider no damping (no external or internal friction). The application of second Newton's law on the mass leads to the well-known ODE of motion:

$$m\ddot{u}(t) + ku(t) = F(t) \quad (1)$$

with $u(t)$ denoting the displacement. Equation (1) is a second order ODE in time $t$, thus it requires two initial conditions:

$$u(0) = u_0, \; \dot{u}(0) = v_0 \quad (2)$$

The standard procedure to solve (1) is to separate the total solution in two parts, a homogeneous solution, $u_h(t)$, and a particular solution, $u_p(t)$, as follows:

$$u(t) = u_h(t) + u_p(t) \quad (3)$$

The first part fulfills the homogeneous ODE:

$$m\ddot{u}_h(t) + ku_h(t) = 0 \quad (4a)$$

or its equivalent:

$$\ddot{u}_h(t) + \omega_e^2 u_h(t) = 0 \quad (4b)$$

where

$$\omega_e^2 = k/m \qquad (5)$$

is the square of the cyclic eigenfrequency (called eigenvalue). The latter corresponds to the free vibration, that is a vibration without external force ($F = 0$). The general solution of (4b) is described by the standard expression

$$u_h(t) = A\cos\omega_e t + B\sin\omega_e t \qquad (6)$$

In general, the particular solution of (1) strongly depends on the algebraic form of the force $F(t)$. Below, two typical loading cases will be studied.

### A. Harmonic excitation

The most popular case is the harmonic force, for example:

$$F(t) = F_0 \sin\omega t \qquad (7)$$

The corresponding particular solution has the same algebraic form, with no phase difference, given by:

$$u_p(t) = U_0 \sin\omega t \qquad (8)$$

Substituting (8) in (1), and considering (7), we obtain:

$$u_p(t) = \frac{F_0}{(\omega_e^2 - \omega^2)} \sin\omega t \qquad (9)$$

Substituting (6) and (9) into (3), and considering also the initial conditions (2), we finally obtain the general solution as a sum of the three following terms:

$$u(t) = u_0 \cos\omega_e t$$
$$+ \frac{1}{\omega_e}\left[v_0 - \frac{\omega F_0}{(\omega_e^2 - \omega^2)}\right]\sin\omega_e t \qquad (10)$$
$$+ \frac{F_0}{(\omega_e^2 - \omega^2)} \sin\omega t$$

Equation (10) depicts that when the cyclic frequency, $\omega$, of the external force [cf. Eq.(7)] approaches the cyclic eigenfrequency, $\omega_e$, of the spring-mass system, the induced elongation of the spring tends to infinity. The tendency of this system to oscillate with larger amplitude at this specific frequency than others is called *resonance*.

Resonance is often observed in the nature, and it is highly related with the damping of the dynamical system. One familiar example is a playground swing, which acts as a pendulum. Pushing a person in a swing in time with the natural interval of the swing (its resonant frequency) will make the swing go higher and higher (maximum amplitude), while attempts to push the swing at a faster or slower tempo will result in smaller arcs. This is because the energy the swing absorbs is maximized when the pushes are 'in phase' with the swing's oscillations, while some of the swing's energy is actually extracted by the opposing force of the pushes when they are not.

Concerning the design of a ($k$, $m$) system, practically we have to tune the proper values of $k$ and $m$, so as the attached machinery that cause the force $F(t)$ operating at a frequency (measured in Hertz [Hz]) should not be close to the eigenfrequency $f_e = \omega_e/(2\pi)$.[1]

### B. Constant force

Another possible type of loading is the case of a constant force, which is sometimes realized in the form of a Heaviside function. The only difference with the above case (A) is that now the particular solution is given by a simpler formula, $u_p(t) = F/k$, so as the general solution of the response is now given by:

$$u(t) = u_0 \cos\omega_e t + \frac{v_0}{\omega_e}\sin\omega_e t$$
$$+ \frac{F}{k}(1 - \cos\omega_e t) \qquad (11)$$

After the above introduction, a question here arises whether the rotation of the spring may cause any remarkable instable phenomena, similar to the aforementioned resonance.

## III. ROTATION

### A. Critical angular frequency

We consider again a spring-mass ($k$, $m$) system, without damping, which now rotates at a constant angular velocity, $\omega_r$, about the end of the spring that was previously fixed (clearly, it does not carry the lumped mass $m$). The axis of rotation is assumed to be perpendicular to the spring. In addition, for the purposes of this paper we assume that the spring appears an infinite resistance to bending deformation.

First of all, the rotation of the ($k$, $m$) system causes a centripetal force on the mass, $m$, which is cancelled by the spring force. Therefore, a trivial application of the aforementioned equilibrium implies that the elongation of the spring, given by the generally false expression:

$$u_{r0} = \frac{m\omega_r^2 R}{k} \qquad (12a)$$

where $R$ is the length of the spring.

It must be pointed out that, Eq(12a) is valid only when the elongation $u_{r0}$ is sufficiently small. It is worthy to mention that, when the angular speed is high, the above approach is quite false. This is because when adding the elongation $u_{r0}$ to the initial radius $R$, the new length becomes $R' = R + u_{r0}$ and now the corresponding centrifugal force will be higher than what Eq(12a) predicts. Therefore, the correct approach is to solve the equation:

$$m\omega_r^2 R' = ku(R' - R), \qquad (12b)$$

from which we obtain:

$$R' = \left(\frac{k}{k - m\omega_r^2}\right) R = \frac{1}{1-\left(\omega_r/\omega_e\right)^2} R \qquad (13)$$

It is worthy to notice that in case that $\omega_r = \omega_e$, Eq(12a) gives the false value $u_{r0} = R$ thus $R' = 2R$ (wrongly), while Eq(13) gives that $R' \to \infty$. Therefore, it is *not* allowable to rotate an elastic rod at an angular frequency $\omega_r$ equal to the resonant one, $\omega_e$. Not only that but it is also not allowable to rotate at a higher angular velocity than $\omega_e$, because in the latter case the length $R'$ would become *negative*! In brief, when a rotation occurs it always must be that $\omega_r < \omega_e$.

### B. Spring elongation and pre-stress

In this section we will deal with the transient phenomenon. As the radius will depend on the time, i.e. it will be $R = R(t)$, we make a minor modification in the symbols. In more details, the initial length of the unloaded spring will be denoted by $R_0$.

For any rotation with an angular frequency $\omega_r < \omega_e$, the spring elongation is correctly given by:

$$u_{r0} = R' - R_0 = \frac{(\omega_r/\omega_e)^2}{1-(\omega_r/\omega_e)^2} R_0$$
$$= \frac{1}{(\omega_e/\omega_r)^2 - 1} R_0 \qquad (14)$$

Therefore, the ground force at the fixation, which is also the center of rotation, is given by ($F_{ground} = F_{spring}$):

$$F_{ground} = ku_{r0} = \frac{kR_0}{(\omega_e/\omega_r)^2 - 1} \qquad (15)$$

### C. Natural frequency

Let us now consider the situation where the spring rotates at a constant angular speed $\omega_r$, at which the spring elongation is $u_{r0}$ and the corresponding (updated) spring length is $R'_0$ given by (13). Here, the subscript '0' was put, so as to denote that this is the initial condition for the subsequent analysis.

In the sequence, in a very slow tempo, we impose –for example– a compressive radial force $F_{int}$, which causes a shortening of the spring, $\Delta u_0$. The latter is related to the given force by the new equilibrium equation:

$$m\omega_r^2 \left(R'_0 - \Delta u_0\right) = k\left(R'_0 - \Delta u_0 - R_0\right) + F_{in}, \qquad (16a)$$

whence

$$F_{int} = \left(m\omega_r^2 - k\right)\left(R'_0 - \Delta u_0\right) + kR_0 \qquad (16b)$$

In (16b), the term $\left(R'_0 - \Delta u_0\right)$ corresponds to the total length of the spring, when the force $F_{int}$ is applied.

When the internal force $F_{int}$ is released (at time $t = 0$), the elongation of the spring will be:

$$U_0 = \left(R'_0 - \Delta u_0\right) - R_0$$
$$= \frac{R_0}{\left(\dfrac{\omega_e}{\omega_r}\right)^2 - 1} - \Delta u_0 \qquad (17)$$

Obviously, the existing difference between the spring force $kU_0$ and the centripetal force $m\omega_r^2 (R_0 + U_0)$, will cause the acceleration of the mass, while for a later time at which the total elongation of the spring will be $U(t)$, the second Newton's law implies:

$$m\ddot{U} + kU = m\omega_r^2 (R_0 + U) \qquad (18)$$

Substituting (17) into (18) the latter becomes:

$$\ddot{U} + \left(\omega_e^2 - \omega_r^2\right)U = \omega_r^2 R_0 \qquad (19)$$

Equation (19) is of major importance. It dictates that the free vibration will occur not at the usual cyclic frequency $\omega_e$, but at:

$$\lambda = \sqrt{\omega_e^2 - \omega_r^2} \qquad (20)$$

This event strongly reminds the superposition of two simultaneous angular velocities (see, for instance, Ref.[1] pp. 247-250), of course with a difference of a minus sign instead of a plus, but it took us considerable effort until we arrive to this conclusion.

### D. Time response

For the sake of briefness and completeness, in the sequence we do not deal with the trivial case of free oscillations, but we consider the general case where after the removal of $F_{int}$, suddenly a generalized force of the form:

$$F(t) = \overline{F} + F_0 \sin \omega t \quad (21)$$

is applied on mass *m*. Obviously, Eq(21) includes both the case of a constant force $\overline{F}$ (including the free vibration $\overline{F} = 0$) and a purely harmonic $F_0 \sin \omega t$.

In this general case, the equation of equilibrium is written as:

$$m\ddot{U} + kU = m\omega_r^2 (R_0 + U) + \overline{F} + F_0 \sin \omega t \quad (22)$$

As previously, we consider the general solution as the sum of the homogeneous solution:

$$U_h(t) = A \cos \lambda t + B \sin \lambda t \quad (23)$$

and also of a particular one. As now the particular solution is a little more complicated, we split it into two parts:

$$U_p(t) = U_{p,1}(t) + U_{p,2}(t) \quad (24)$$

where they fulfill the ODEs:

$$\ddot{U}_{p,1} + \lambda^2 U_{p,1} = \omega_r^2 R_0 + \overline{F}/m \quad (25a)$$

$$\ddot{U}_{p,2} + \lambda^2 U_{p,2} = F_0 \sin \omega t / m \quad (25b)$$

It can be easily found that:

$$U_{p,1} = \frac{\omega_r^2 R_0 + \overline{F}/m}{\lambda^2}$$

$$U_{p,2} = \frac{F_0}{m(\lambda^2 - \omega^2)} \sin \omega t \quad (26)$$

Considering the initial conditions

$$U(0) = U_0, \ \dot{U}(0) = 0 \quad (27)$$

the general solution consists of three harmonic terms as follows:

$$U(t) = U_0 \cos \lambda t$$
$$+ \frac{(\omega_r^2 R_0 + \overline{F}/m)}{\lambda^2}(1 - \cos \lambda t) \quad (28)$$
$$+ \frac{F_0}{m(\lambda^2 - \omega^2)}\left(\sin \omega t - \frac{\omega}{\lambda}\sin \lambda t\right)$$

Obviously, none of the two denominators in (28) are allowed to vanish. This means that:

1) The angular speed $\omega_r$ must be always smaller than $\omega_e$ ($\lambda \neq 0$).
2) When $\lambda \neq 0$ as above, the angular frequency $\omega$ of the external force $F(t)$ must never become equal to $\lambda = \sqrt{\omega_e^2 - \omega_r^2}$.

Finally, it is remarkable that the ground force $F_{ground} = kU(t)$ is a *harmonic* function.

## IV. AN IMAGINARY WORLD WHERE THE NATURAL FREQUENCY DOES NOT DEPEND ON ROTATION

If natural frequency was not dependent on the rotation, many unusual phenomena could happen.

### A. THE LUMPED SYSTEM

Actually, let us now assume that the external force, *F*, has a constant value and the spring is pre-stressed due to the angular speed $\omega_r$. Here, one student could wrongly make the assumption that the aforementioned pre-stressed field is a kind of body-forces that remain unaltered during a subsequent loading by the force *F*, which causes an additional displacement $u_r(t)$ on the mass *m*. With respect to the pre-stressed spring, it is quite reasonable to assume vanishing initial conditions ($u_0 = 0, v_0 = 0$). A rotating observer understands that a constant force *F* is exerted on the (*k*, *m*) system, thus the finding of Eq(11) can be used and, therefore, the additional elongation (or shortening) of the spring is now given as:

$$u_r(t) = \frac{F}{k}(1 - \cos \omega_e t) \quad (29)$$

Again, the mistake in (29) is that $\omega_e$ had to be replaced by $\lambda = \sqrt{\omega_e^2 - \omega_r^2}$. In this context where the force *F* is constant, the only possible *synchronization* might occur combing the angular velocity, $\omega_r$, with the hypothetically constant cyclic eigenfrequency, $\omega_e$, of the spring-mass system.

Due to the length change in the spring given by (29), the ground force equals to the product *ku*. Therefore, the ground force, which lies in the radial direction along the spring, would be given by:

$$\vec{F}_{ground}^{radial}(t) = \begin{Bmatrix} F_{rx} \\ F_{ry} \end{Bmatrix} = F(1 - \cos \omega_e t) \cdot \vec{e}_r, \quad (30)$$

where $\vec{e}_r = (\cos\theta, \sin\theta)^T$ denotes the rotating radial unit vector. The polar angle, $\theta$, is given in terms of its initial value, $\theta_0$, by:

$$\theta = \theta_0 + \omega_r t \qquad (31)$$

It is remarkable that as a rotating observer sees a *moving* mass, $m$, he/she has to add a *Coriolis* force, $-2m\vec{\omega}_r \times \vec{u}_r$, perpendicularly to the line of the spring and directed towards the opposite of the tangential unit vector $\vec{e}_\theta$. This perpendicularity, in conjunction with the assumption of infinite resistance to bending, allows for superimposing the axial vibration and the additional action of Coriolis force. Then, the Coriolis force is transferred, as a shear force, to the ground:

$$\vec{F}_{ground}^{shear}(t) = \begin{Bmatrix} F_{cx} \\ F_{cy} \end{Bmatrix} = -2\left(\frac{\omega_r}{\omega_e}\right) F \sin\omega_e t \cdot \vec{e}_\theta, \qquad (32)$$

with $\vec{e}_\theta = (-\sin\theta, \cos\theta)^T$.

As a result, the vertical component of the total ground force is given by:

$$\vec{F}_{ground}^{total}(t) = \vec{F}_{ground}^{radial}(t) + \vec{F}_{ground}^{shear}(t), \qquad (33)$$

For reasons that will be explained in the next subsection B, the interesting point (if the assumption of independency was correct) here is the sign of the components of the abovementioned force $\vec{F}_{ground}^{total}$ and the way its *impulse* varies with the time. Below, in subsection B (**Fig. 3**), we will see that under certain conditions we could achieve a desired sign, for example, positive value, of the vertical ground force component.

According to standard literature, Ref.(1, p. 212), "the integral of a force over the time interval during which the force acts is called the impulse **J** of the force". It is well known that the change in linear momentum, $\mathbf{p}_f - \mathbf{p}_i$, equals to the aforementioned impulse:

$$\mathbf{J} = \int_{t_i}^{t_f} \mathbf{F} dt. \qquad (34)$$

The corresponding impulses of the two abovementioned inertial forces $\vec{F}_{ground}^{radial}$ and $\vec{F}_{ground}^{shear}$ are denoted by the vectors:

$$\mathbf{J}_{ground}^{radial} = \{J_{rx} \quad J_{ry}\}^T \qquad (18a)$$

and

$$\mathbf{J}_{ground}^{Coriolis} = \{J_{cx} \quad J_{cy}\}^T, \qquad (18b)$$

Again, if our assumption were correct, the components of the abovementioned vectors can be described by multiply defined analytical expressions, which depend on the type of synchronization, i.e. if the angular velocity $\omega_r$ of the rotation coincides with the cyclic eigenfrequency $\omega_e$ or not, as follows.

**B1. When $\omega_r \neq \omega_e$**

*Ground forces:*

$$\left(F_{ground}^{total}\right)_x = F\left[-\cos\theta(\cos\omega_e t - 1) + 2\left(\frac{\omega_r}{\omega_e}\right)\sin\theta\sin\omega_e t\right]$$

$$\left(F_{ground}^{total}\right)_y = F\left[-\sin\theta(\cos\omega_e t - 1) - 2\left(\frac{\omega_r}{\omega_e}\right)\cos\theta\sin\omega_e t\right]$$

$$(36)$$

*Impulse:*

$$J_{rx} = F\left[\frac{\sin\theta}{\omega_r} + \frac{-\omega_e \cos\theta\sin\omega_e t + \omega_r \sin\theta\cos\omega_e t}{(\omega_e^2 - \omega_r^2)}\right]$$

$$J_{ry} = F\left[\frac{-\cos\theta}{\omega_r} + \frac{-\omega_e \sin\theta\sin\omega_e t - \omega_r \cos\theta\cos\omega_e t}{(\omega_e^2 - \omega_r^2)}\right]$$

$$(37)$$

and

$$J_{cx} = F\frac{-2\omega_r(\omega_e \sin\theta\cos\omega_e t - \omega_r \cos\theta\sin\omega_e t)}{\omega_e(\omega_e^2 - \omega_r^2)}$$

$$J_{cy} = F\frac{2\omega_r(\omega_e \cos\theta\cos\omega_e t + \omega_r \sin\theta\sin\omega_e t)}{\omega_e(\omega_e^2 - \omega_r^2)}$$

$$(38)$$

**B2. When $\omega_r = \omega_e$**

*Ground forces:*

$$\left(F_{ground}^{total}\right)_x = F\left[2\sin\theta\sin\omega_e t - \cos\theta(\cos\omega_e t - 1)\right]$$

$$\left(F_{ground}^{total}\right)_y = \frac{F}{2}\left[2\sin\theta - 3\sin(\theta_0 + 2\omega_e t) + \sin\theta_0\right]$$

$$(39)$$

*Impulse:*

$$J_{rx} = F\left[-\left(\frac{\cos\theta_0}{2}\right)t + \frac{4\sin\theta - \sin(\theta_0 + 2\omega_e t)}{4\omega_e}\right]$$

$$J_{ry} = F\left[-\left(\frac{\sin\theta_0}{2}\right)t - \frac{4\cos\theta - \cos(\theta_0 + 2\omega_e t)}{4\omega_e}\right]$$

$$(40)$$

and

$$J_{cx} = F\left[(\cos\theta_0)t - \frac{\sin(\theta_0 + 2\omega_e t)}{2\omega_e}\right]$$

$$J_{cy} = F\left[(\sin\theta_0)t + \frac{\cos(\theta_0 + 2\omega_e t)}{2\omega_e}\right]$$

$$(41)$$

In other words, while in resonance at $\omega_r = \omega_e$ (wrong hypothesis) we find the response to tend to infinity, in the present synchronization condition the equations would jump to a smooth state but they would appear a term that is linearly proportional to the time *t*. When the reader reads the subsection B, he/she will better understand why this term had to be anticipated.

Collecting the radial and the Coriolis forces for the case where the rotation is synchronized with the axial free vibration ($\omega_r = \omega_e$), the sum is given by the following vector:

$$\mathbf{J}_{ground}^{total} = F \left\{ \begin{array}{l} \left(\dfrac{\cos\theta_0}{2}\right)t + \dfrac{4\sin\theta - 3\sin(\theta_0 + 2\omega_e t)}{4\omega_e} \\ \left(\dfrac{\sin\theta_0}{2}\right)t + \dfrac{-4\cos\theta + 3\cos(\theta_0 + 2\omega_e t)}{4\omega_e} \end{array} \right\}$$
(42)

Therefore, both impulse components finally would appear a linear term with the time *t* (if the hypothesis were correct). It is worth-mentioning that if we were interested in maximizing the upward (vertical, i.e. the *y*-) component, the best choice would be the value $\theta_0 = \pi/2$.

Concerning the reaction of the force *F*, which in a closed system should be transferred to the ground, it is worth-mentioning that its impulse could not cancel the linear terms appearing in Eq(42). Thus mechanical antigravity would have been achieved.

## B. THE CONTINUUM MODEL

Moreover, the previous subsection A concerning the synchronization was inspired during a practical concerning programming of the numerical solution for wave propagation in elastic rods. This example concerns an elastic bar of length *R*, which is fixed at its left end, while the right end is loaded by a force of Heaviside type. While in the one-degree vibrating system (dealt in subsection A) the force is immediately transferred to the ground, in the real continuum it takes some time ($T = R/c$) until the wave reaches the fixed end. For given elastic modulus *E* and mass density $\rho$, it holds that $c = \sqrt{E/\rho}$.[1] The matter of impact has been dealt in a general manner by Whittaker,[4] while for a complete analysis of impulse loading of a rod we refer to Graff[5] as well as to Pipes and Harville.[6] Also, the wave reflection has been dealt in an instructive way in a very recent paper.[7] Upon the arrival of the incident wave at the left (fixed) end the displacement of the right end obtains the static solution dictated by Hooke's law ($\sigma = E\varepsilon = E u_{static}/R$). Immediately after the deflection, the displacement of the right end continues to increase until the wave is reflected at the right end, at time 2*T*; then the right end is displaced twice as the static motion ($u_{dynamic} = 2u_{static}$). It is useful to make the students aware that in many engineering practices, such as in mechanical engineering, this is a reason that professional manuals suggest to use an additional (dynamic) safety factor equal to two.

The conclusion of the above analysis is that the ground force at the fixed left end becomes twice than the load at the right end but the duration is half the exact theoretical period $T_e = 4R/c$, as shown in **Fig. 1**. It is worth-mentioning that the period calculated by the (*k*, *m*) lumped system appears an error of about 11 percent but it allows for an analytical handling of the entire problem discussed in Section III. The graph in **Fig. 1** could be the initiation point to think that a rotation owing the same period with $T_e$ would be capable of blocking the direction of the ground force in the desired direction.

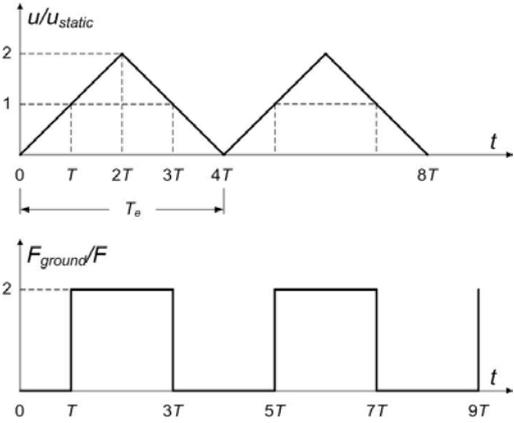

Fig. 1. Variation of the longitudinal displacement at the loaded end of the rod (top) and the ground force reaction (bottom).

In more details, let us assume that at time *t*=0 the rod is at the vertical position with the loaded end at the lower level. Under the condition $\omega_r = \omega_e$, at time *t* = *T* the wave front will arrive at the point O when the rod will have obtained its horizontal position OE' shown by the dashed line in **Fig. 2**. Obviously, for the entire time interval [*T*,3*T*] the rod will be found above the horizontal line, and it will exert *on the ground* a vertical component given by:

$$F_z = 2F \sin\theta \qquad (43)$$

Therefore, the object to which the rotating bar is attached will undertake the force given by Equation (43), which is *non-negative* as shown in **Fig. 3**. This correct conclusion gives the impression that the ground force always is upward.

In case that $\omega_r \neq \omega_e$, the student could continue with a homework in which the accurate Coriolis forces of the continuum would be considered.

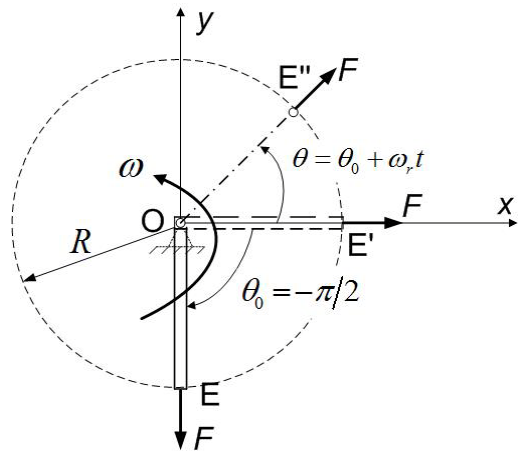

Fig. 2. An elastic rod of length *R* is subjected to a constant load *F* and rotates at a constant angular velocity. At time *t* = 0 it would be found at the position OE, while at *t* = *T* = $T_e$/4 it would be found at the horizontal position OE'.

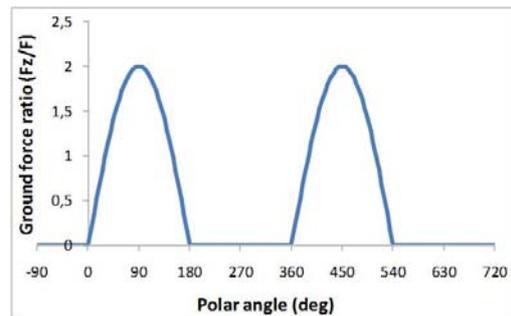

Fig. 3. Hypothetical lift of the ground force (vertical component) due to the synchronization.

## V. CONCLUSION

In this study, we presented the subject of "resonance", starting from the usual conditions of translational motion and continuing with rotation. Based on the simplest possible background of a linear spring without damping connected to a lumped mass, we concluded that rotation highly influences the natural frequency. In order to please the reader, and taking causation from several internet references that *rotation* makes things (in mechanics) different, we presented an imaginary world in which the natural frequency does not depend on the rotation. The major conclusion is that, under certain circumstances, in that virtual world it would be possible to create non-harmonic ground/spring forces, a hypothetical result that is closely related to the achievement of mechanical antigravity.